\documentclass[copyright]{eptcs}
\providecommand{\event}{TTC 2011} 

\usepackage{tikz}

\usepackage{listings}
\usepackage{graphicx}
\usepackage{xspace}
\usepackage{csquotes}

\newcommand{\origttfamily}{}
\let\origttfamily=\ttfamily
\renewcommand{\ttfamily}{\origttfamily \hyphenchar\font=`\-}

\newcommand{\name}[1]{{\sc #1}\xspace}
\newcommand{\grgen}{\name{GrGen}}
\newcommand{\grgennet}{\name{GrGen.NET}}
\newcommand{\grshell}{\name{GrShell}}

\lstdefinelanguage{grgen}
{morekeywords={rule,test,pattern,replace,if, iterated, eval,negative,node,edge,class,model,connect,extends,using,return,typeof,abstract,modify,delete,enum,int,string,double,float,boolean,alternative,emit,exec},
sensitive=true,
morecomment=[l]{//},
morecomment=[s]{/*}{*/},
morestring=[b]",
tabsize=2,
basicstyle=\ttfamily\footnotesize,
keywordstyle=\itshape,
}

\lstdefinelanguage{xgrs}
{morekeywords={xgrs,debug,import,dump,add,set,node,edge,exit,graph,group,by,hidden,new,outgoing,labels,off,shortinfotag,exclude,shape,show,color,rhomb,white},
sensitive=true,
morecomment=[l]{\#},
morecomment=[s]{/*}{*/},
morestring=[b]",
basicstyle=\ttfamily\footnotesize,
keywordstyle=\itshape,
}

\title{Solving the TTC 2011 Compiler Optimization Case with GrGen.NET}
\author{Sebastian Buchwald \quad \quad Edgar Jakumeit
\institute{Karlsruhe Institute of Technology (KIT)}
\email{buchwald@kit.edu \quad \quad \phantom{~~~~~~~~~~~~~~~~}}
}
\def\titlerunning{Solving the TTC 2011 Compiler Optimization Case with GrGen.NET}
\def\authorrunning{Sebastian Buchwald and Edgar Jakumeit}
\begin{document}
\maketitle

\begin{abstract}
The challenge of the Compiler Optimization Case \cite{compileroptimizationcase} is to perform local optimizations and instruction selection on the graph-based intermediate representation of a compiler.
The case is designed to compare participating tools regarding their performance.
We tackle this task employing the general purpose graph rewrite system \grgennet (\url{www.grgen.net}).
\end{abstract}

\section{What is GrGen.NET?}

\grgennet\ is an application domain neutral graph rewrite system, which started as a helper tool for compiler construction.
It is well suited for solving this challenge due to its origin, thus our solution can be seen as a reference, even though the origins were left behind a while ago.
The feature highlights regarding practical relevance are:
\begin{description}
\item[Fully Featured Meta Model:] \grgennet\ uses attributed and typed multigraphs with multiple inheritance on node/edge types. Attributes may be typed with one of several basic types, user defined enums, or generic set, map, and array types.
\item[Expressive Rules, Fast Execution:] The expressive and easy to learn rule specification language allows straightforward formulation of even complex problems, with an optimized implementation yielding high execution speed at modest memory consumption.
\item[Programmed Rule Application:] \grgennet\ supports a high-level rule application control language, Graph Rewrite Sequences (GRS), offering sequential, logical, iterative and recursive control plus variables and storages for the communication of processing locations between rules.
\item[Graphical Debugging:] \grshell, \grgennet's command line shell, offers interactive execution of rules, visualising together with yComp the current graph and the rewrite process. This way you can see what the graph looks like at a given step of a complex transformation and develop the next step accordingly. Or you can debug your rules and sequences.
\item[Extensive User Manual:] The \grgennet User Manual \cite{GrGenUserManual} guides you through the various features of \grgennet, including a step-by-step example for a quick start.
\end{description}

\section{Constant Folding}
The first task is to perform constant folding.
Constant folding transforms operations which have only \texttt{Const} operands into a \texttt{Const} itself, e.g.\ to transform $1 + 2$ into $3$.
This is a local optimization requiring only rules referencing local graph context.

\subsection{Driver and Data Flow}

Constant folding is carried out from the driver sequence given in \autoref{fig:driversequence.grsi},
which performs inside a main loop in each step
i) first constant folding along data flow with a wavefront algorithm,
and then ii) control flow folding together with clean-up tasks.

Graph rewrite sequences are the rule application control language of \grgen.
The most fundamental operation is a rule application denoted by the rule name, with parameters given in parenthesis, and optionally an assignment of output values to variables (given in parenthesis, too).
By using all bracketing \verb#[r]# we execute the rule \texttt{r} for all matches in the host graph.
The then-right operator \texttt{;>} executes the left sequence and then the right sequence, returning as result of execution the result of the execution of the right sequence.
The potential results of sequence execution are \emph{success} equaling \texttt{true} and \emph{failure} equaling \texttt{false}; a rule which matches (at least once) counts as success.
With the postfix star \texttt{*} we iterate the preceding sequence as long as it succeeds.
The result of a star iteration is always success, in contrast to the plus \texttt{+} postfix which requires the preceding sequence to match at least once in order to succeed; so a sequence of rules with plus postfixes linked by strict disjunction operators \verb#|# succeeds (\texttt{true}) if one of the rules matches.
Conjunction \verb#&# is available as well, so are the lazy versions \verb#&&# and \verb#||# of the operators not executing the right sequence in case the result of the left sequence already determines the outcome.
The prefix exclamation mark operator negates the result of sequence execution, in \autoref{fig:driversequence.grsi} we negate the
value from variable \texttt{isEmpty}.
The backslashes \verb#\# allow to concatenate multiple lines into one.

\begin{figure}[h]
\begin{lstlisting}[language=xgrs]
xgrs now:set<FirmNode>=set<FirmNode>{} ;> next:set<FirmNode>=set<FirmNode>{} ;>\
    ( [collectConstUsers(dummy, now)] ;>\
        (wavefront(now, next) ;> now.clear() ;> tmp=now ;> now=next\
            ;> next=tmp ;> isEmpty=now.empty() ;> !isEmpty)* ;>\
        [foldAssociativeAndCommutative] ;>\
        (foldCond+ | removeUnreachablePhiOperand+ |\
            removeUnreachableBlock+ | removeUnreachableNode+)* ;>\
        fixEdgePosition* ;>\
        (mergeConsts+ | removeUnusedNode+ | mergeBlocks+)\
    )*
\end{lstlisting}
	\caption{Driver sequence for constant folding.}
	\label{fig:driversequence.grsi}
\end{figure}

\noindent In the driver sequence two empty storage sets are created initially for the wavefront algorithm: \texttt{now} which will contain the users of constant nodes which are visited in this step and \texttt{next} which will contain the users of constant nodes which will be visited in the next step.
Then the main loop is entered.
There the rule \texttt{collectConstUsers} is executed on all available matches, collecting all users of \texttt{Const} nodes available in the graph into the storage set \texttt{now} (the \texttt{dummy} variable was never assigned and the \texttt{collectConstUsers} rule was specified to search for a valid entity if the first parameter is undefined).
This is the initial stone thrown into the water of our program graph.
From it on a wavefront is iterated following the data flow edges until it comes to a halt because no new constants were created anymore.
A wavefront step is encapsulated in the subsequence \texttt{wavefront} given in \autoref{fig:wavefront.grsi};
it visits all operations of its first argument,
and adds the operations which are users of the constants it created newly with constant folding into the set given as its second argument.
After this subsequence was called on the \texttt{now} argument the \texttt{now} and \texttt{next} sets are swapped (the set referenced by \texttt{now} is cleared, \texttt{now} is assigned the set referenced by the \texttt{next}, and next is assigned the empty set previously pointed to by \texttt{now}).
The wavefront iteration is stopped when the \texttt{now} set of the next iteration step is empty.

\begin{figure}[ht]
\begin{lstlisting}[language=xgrs]
def wavefront(constUsersNow:set<FirmNode>,\
              constUsersNext:set<FirmNode>) {\
	for{\
		cu:FirmNode in constUsersNow;\
		((c)=foldNot(cu) || (c)=foldBinary(cu) ||\
		 (c)=foldPhi(cu) & false\
		)\
		&& [collectConstUsers(c, constUsersNext)]\
	} ;> constUsersNow.clear()\
}
\end{lstlisting}
	\caption{Wavefront step for constant folding.}
	\label{fig:wavefront.grsi}
\end{figure}

\noindent A wavefront step defined by the subsequence given in \autoref{fig:wavefront.grsi}
iterates with a \texttt{for} loop over the users of constants contained in the storage set parameter \texttt{constUsersNow},
binding them to an iteration variable \texttt{cu} of type \texttt{FirmNode}.
This constant is used as input to the rules \texttt{foldNot}, \texttt{foldBinary} and \texttt{foldPhi} really doing the folding in case all of the arguments to the operations they handle are constant.
They return the newly created constant by folding, the rule \texttt{collectConstUsers} then adds all users of this constant to the \texttt{constUsersNext} storage set.

Before we take a closer look at the rules, let us start with a short introduction into the syntax of the rule language:
Rules in GrGen consist of a pattern part specifying the graph pattern to match and a nested rewrite part specifying the changes to be made.
The pattern part is built up of node and edge declarations or references with an intuitive syntax:
Nodes are declared by \texttt{n:t}, where \texttt{n} is an optional node identifier, and \texttt{t} its type.
An edge \texttt{e} with source \texttt{x} and target \texttt{y} is declared by \texttt{x -e:t-> y}, whereas \texttt{-->} introduces an anonymous edge of type \texttt{Edge}.
Nodes and edges are referenced outside their declaration by \texttt{n} and \texttt{-e->}, respectively.
Attribute conditions can be given within \texttt{if}-clauses.

The rewrite part is specified by a \texttt{replace} or \texttt{modify} block nested within the rule.
With \texttt{replace}-mode, graph elements which are referenced within the replace-block are kept, graph elements declared in the replace-block are created, and graph elements declared in the pattern, not referenced in the replace-part are deleted.
With \texttt{modify}-mode, all graph elements are kept, unless they are specified to be deleted within a \texttt{delete()}-statement.
Attribute recalculations can be given within an \texttt{eval}-statement.
In addition to rewriting, \grgen supports relabeling, we prefer to call it retyping though.
Retyping is specified with the syntax \texttt{y:t<x>}: this defines \texttt{y} to be a retyped version of the original node \texttt{x}, retyped to the new type \texttt{t}; for edges the syntax is \texttt{-y:t<x>->}.

These and the language elements we introduce later on are described in more detail in our solution of the Hello World case \cite{helloworldsolutiongrgennet},
and especially in the extensive GrGen.NET user manual \cite{GrGenUserManual}.

\autoref{fig:foldBinary.grg} shows the constant folding rule for \texttt{Binary} operations.
It takes the node with the \texttt{Const} operand as parameter and returns the folded \texttt{Const}.
The rule matches both \texttt{Const} operands and has an \texttt{alternative} statement that contains one case for each binary operation.
Within the cases the value of the new \texttt{Const} is computed.
In almost all cases the computation is as simple as in the \texttt{foldAdd} case.
The only exception is the \texttt{foldCmp} case that also needs to consider the \texttt{relation} of the \texttt{Cmp} to compute the resulting value of the \texttt{Const}.
Finally, at the end of the rule execution, the \texttt{exec} statement relinks the users of the \texttt{Binary} to the newly created \texttt{Const} and deletes the \texttt{Binary} from the graph (by \texttt{exec}uting the given sequence).
The constant folding for unary \texttt{Not} nodes is straight forward and not further discussed;
$n$-ary \texttt{Phi} nodes are of interest again.
\autoref{fig:foldPhi.grg} shows the corresponding rule.
It first matches the \texttt{Phi} node and one \texttt{Const} operand and then iteratively edges to the same operand and to the \texttt{Phi} itself.
If this covers all edges we can replace the \texttt{Phi} with the \texttt{Const} operand.

\subsection{Control Flow and Cleanup}

Let us mentally return to the driver sequence in \autoref{fig:driversequence.grsi};
after the wavefront which folded alongside data flow has collapsed,
execution continues with folding condition nodes at the interface of data flow to control flow
and with folding control flow proper (jumps and blocks).
In addition there are some clean up task left to be executed.
If no control flow was folded leading to further data flow folding possibilities, the main loop is left.

The rule shown in \autoref{fig:foldCond.grg} is responsible for folding \texttt{Cond}itional jumps.
Depending on the value of the constant,
either the edge of type \texttt{True} gets deleted and the edge of type \texttt{False} retyped to an edge of type \texttt{Controlflow}, or the edge of type \texttt{False} gets deleted and the edge of type \texttt{True} retyped to an edge of type \texttt{Controlflow}.
Since there is only one jump target left, we also retype the conditional jump to a simple jump of type \texttt{Jmp}.
Due to the folding of condition jumps, there may be unreachable blocks which can be deleted.
We use two rules to remove unreachable code: the rule shown in \autoref{fig:removeUnreachableBlock.grg} deletes unreachable \texttt{Block}s, the rule shown in \autoref{fig:removeUnreachableNode.grg} deletes nodes without a \texttt{Block}.
Furthermore, we also need to adapt \texttt{Phi} nodes if they have an operand without a \texttt{Controlflow} counterpart in the \texttt{Block}.
\autoref{fig:removeUnreachablePhiOperand.grg} shows the corresponding rule that matches a \texttt{Phi} and deletes an operand that has no \texttt{Controlflow} counterpart.
After the deletion we fix the position of all edges using the \texttt{fixEdgePosition} rule.

The solution contains two rules \texttt{removeUnusedNode} and \texttt{mergeBlocks} that simplify the graph.
The first rule removes a node that is not used, i.e.\ a node which has no incoming edges.
This rule is similar to the \texttt{removeUnreachableBlock} rule.
If there is a \texttt{Block} \texttt{block1} with only one outgoing edge of type \texttt{Controlflow} and this edge leads to a node of type \texttt{Jmp} contained in \texttt{Block} \texttt{block2}, then we remove the \texttt{Jmp} and merge \texttt{block1} and \texttt{block2} with the second rule.
This rule is able to remove the chain of \enquote{empty} \texttt{Block}s that occurs in the running example of the case description.

The verifier mentioned in the case descpription was implemented with the tests in \texttt{Verifier.gri}, called from a subsequence \texttt{verify} defined in \texttt{Verifier.grsi}; the subsequence is used with the statement \texttt{validate xgrs verify} before and after the transformations checking the integrity of the graph.

\subsection{Folding More Constants}
\autoref{fig:foldAssociativeAndCommutative.grg} shows a rule that folds constants for the term $(x\ast c1) \ast c2$ where $\ast$ is a associative and commutative operation.
This rule enables us to solve the test cases provided by the GReTL solution.
In contrast to the GReTL rule, this rule does not change the structure of the graph.

\section{Instruction Selection}

The instruction selection task transforms the intermediate representation (IR) into a target-dependent representation (TR). 
It can be considered as some kind of model transformation.
The TR supports immediate instructions, i.e.\ a \texttt{Const} operand can be encoded within the instruction.
Our solution consists of one rule per target instruction which means we have at most two rules for each IR operation: one for the immediate variant and one for the variant without immediate.

Before we start matching all immediate operations, we apply the auxiliary rule shown in \autoref{fig:normalizeConst.grg}.
It matches commutative operations with a \texttt{Const} operand at \texttt{position} $0$, but not at \texttt{position} $1$, and then exchanges these operands to ensure that the constant operand is at \texttt{position} $1$.
Thus, the rule ensures deterministic behaviour in case of two constant operands.

\autoref{fig:iselAddI.grg} shows exemplarily the rule that creates an \texttt{AddI} operation.
It retypes the \texttt{Add} to an \texttt{TargetAddI}, deletes the \texttt{Dataflow} edge to the constant and stores the value of the \texttt{Const} node in the \texttt{value} attribute.
Since the other rules for immediate operations are very similar, we omit them here.
Instead we want to introduce the rule for creating \texttt{TargetLoad}s given in \autoref{fig:iselLoad.grg}.
Again, we retype the original operation to the corresponding target operation.
Since \texttt{Load} is a memory operation, we also need to set the \texttt{volatile} attribute.

The instruction selection rules are applied by the sequence shown in \autoref{fig:isel.grs}.
The \texttt{[rule]} operator matches and rewrites all occurrences of the given \texttt{rule}.
Hence, we first create all immediate operations, and then the non-immediate operations for the remaining nodes.

\section{Conclusion}

We presented a \grgen solution for both tasks, constant folding as well as instruction selection, covering all test cases.
The solution employing pure graph rewriting is pretty concise for the constant folding task but admittedly less so for the instruction selection task,
which requires a good deal of (very simple) graph relabeling rules, one rule per operation plus a further rule for the operations with immediate.
The conciseness could be improved by stepping an abstraction level up by implementing a generator emitting GrGen code, which is what we did in~\cite{Buchwald2010} and~\cite{SG:07} and what we recommend for the classes of applications which would lead to repetitive code.

\begin{figure}
\centering
\includegraphics[scale=0.38]{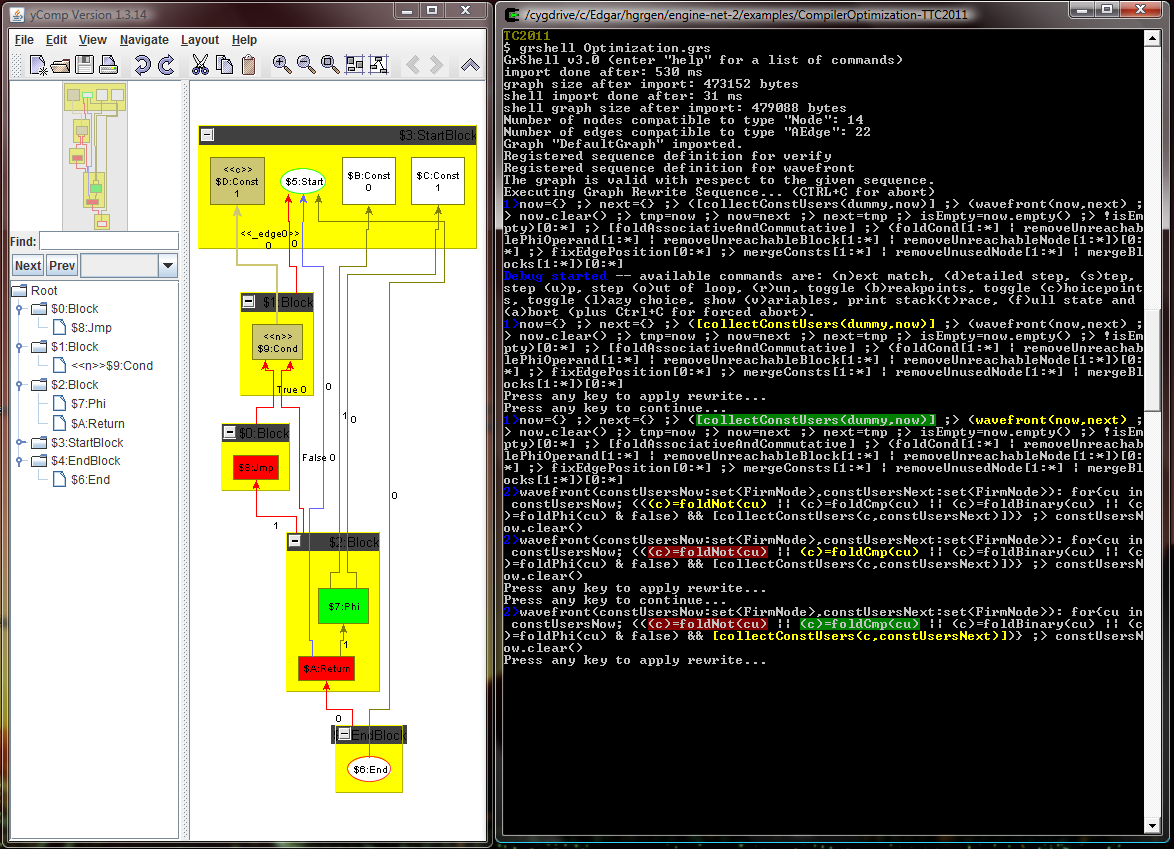}
\caption{A situation from constant folding}
\label{fig:OptimizationSituation}
\end{figure}

The debugger included in the GrShell, which allows to execute the sequences step-by-step under user control, was a great help during development of the optimizations.
Especially its graph visualizations, which are depicted in \autoref{fig:OptimizationSituation} and in the case description \cite{compileroptimizationcase}.

Due to the wavefront algorithm only visiting relevant operations and the core speed of the generated code we were able to give a solution with excellent performance.
For the largest GReTL test case -- consisting of 27993 nodes and 55981 edges -- the execution speed is 6.3 seconds for constant folding and 111 milliseconds for instruction selection.
For the largest test shipped with this case -- consisting of 277529 nodes and 824154 edges -- we need 11 seconds for constant folding and 1.2 seconds for instruction selection.
This is more than one order of magnitude better than the next-closest competitor.
The measurements were made using Ubuntu 11.04 and Mono 2.6.7 on a Core2Duo E6550 with 2.33 GHz.
The numbers are even better for the SHARE machine(s) hosting the GrGen image \cite{share} at the time of writing.

Our solution does not apply rules in parallel.
However, we consider it as a benchmark for parallel solutions.
Unfortunately, nobody submitted a parallel solution, which was the main purpose of publishing the challenge: we were hoping for hints about the potential benefits of parallelization.
So the question which speed up can be gained by parallel rule application is still open.

\nocite{*}
\bibliographystyle{eptcs}
\bibliography{refs}

\begin{thebibliography}{1}
\providecommand{\bibitemdeclare}[2]{}
\providecommand{\urlprefix}{Available at }
\providecommand{\url}[1]{\texttt{#1}}
\providecommand{\href}[2]{\texttt{#2}}
\providecommand{\urlalt}[2]{\href{#1}{#2}}
\providecommand{\doi}[1]{doi:\urlalt{http://dx.doi.org/#1}{#1}}
\providecommand{\bibinfo}[2]{#2}

\bibitemdeclare{misc}{GrGenUserManual}
\bibitem{GrGenUserManual}
\bibinfo{author}{Jakob Blomer}, \bibinfo{author}{Rubino Gei{\ss}} \&
  \bibinfo{author}{Edgar Jakumeit} (\bibinfo{year}{2011}):
  \emph{\bibinfo{title}{{The GrGen.NET User Manual}}}.
\newblock \bibinfo{howpublished}{\url{http://www.grgen.net}}.

\bibitemdeclare{inproceedings}{compileroptimizationcase}
\bibitem{compileroptimizationcase}
\bibinfo{author}{Sebastian Buchwald} \& \bibinfo{author}{Edgar Jakumeit}
  (\bibinfo{year}{2011}): \emph{\bibinfo{title}{{Compiler Optimization: A Case
  for the Transformation Tool Contest}}}.
\newblock In \bibinfo{editor}{\bibinfo{editor}{{Van Gorp}}} et~al.
  \cite{ttc2011eptcs}.

\bibitemdeclare{inproceedings}{helloworldsolutiongrgennet}
\bibitem{helloworldsolutiongrgennet}
\bibinfo{author}{Sebastian Buchwald} \& \bibinfo{author}{Edgar Jakumeit}
  (\bibinfo{year}{2011}): \emph{\bibinfo{title}{Saying Hello World with
  {GrGen}.{NET} -- A Solution to the {TTC} 2011 Instructive Case}}.
\newblock In \bibinfo{editor}{\bibinfo{editor}{{Van Gorp}}} et~al.
  \cite{ttc2011eptcs}.

\bibitemdeclare{inproceedings}{Buchwald2010}
\bibitem{Buchwald2010}
\bibinfo{author}{Sebastian Buchwald} \& \bibinfo{author}{Andreas Zwinkau}
  (\bibinfo{year}{2010}): \emph{\bibinfo{title}{Instruction Selection by Graph
  Transformation}}.
\newblock In: {\sl \bibinfo{booktitle}{Proceedings of the 2010 international
  conference on Compilers, architectures and synthesis for embedded systems}},
  \bibinfo{series}{CASES '10}, \bibinfo{publisher}{ACM}, \bibinfo{address}{New
  York, NY, USA}, pp. \bibinfo{pages}{31--40}, \doi{10.1145/1878921.1878926}.

\bibitemdeclare{misc}{share}
\bibitem{share}
\bibinfo{author}{Edgar Jakumeit} \& \bibinfo{author}{Sebastian Buchwald}
  (\bibinfo{year}{2011}): \emph{\bibinfo{title}{{SHARE} demo related to the
  paper Solving the {TTC} 2011 Compiler Optimization Case with {GrGen}.{NET}}}.
\newblock
  \bibinfo{howpublished}{\url{http://is.ieis.tue.nl/staff/pvgorp/share/?page=C%
onfigureNewSession&vdi=XP-TUe_TTC11_GrGen_v2.vdi}}.

\bibitemdeclare{conference}{SG:07}
\bibitem{SG:07}
\bibinfo{author}{Andreas Sch{\"o}sser} \& \bibinfo{author}{Rubino Gei{\ss}}
  (\bibinfo{year}{2008}): \emph{\bibinfo{title}{Graph Rewriting for Hardware
  Dependent Program Optimizations}}.
\newblock In \bibinfo{editor}{A.~Sch{\"u}rr}, \bibinfo{editor}{M.~Nagl} \&
  \bibinfo{editor}{A.~Z{\"u}ndorf}, editors: {\sl \bibinfo{booktitle}{Proc. 3rd
  Intl. Workshop on Applications of Graph Transformation with Industrial
  Relevance (AGTIVE '07)}}, \bibinfo{series}{LNCS},
  \bibinfo{publisher}{Springer}, \doi{10.1007/978-3-540-89020-1\_17}.

\bibitemdeclare{proceedings}{ttc2011eptcs}
\bibitem{ttc2011eptcs}
\bibinfo{editor}{Pieter {Van Gorp}}, \bibinfo{editor}{Steffen Mazanek} \&
  \bibinfo{editor}{Louis Rose}, editors (\bibinfo{year}{2011}):
  \emph{\bibinfo{title}{{TTC} 2011: Fifth Transformation Tool Contest,
  Z\"urich, Switzerland, June 29-30 2011, Post-Proceedings}}.
  \bibinfo{publisher}{{EPTCS}}.

\end{thebibliography}

\appendix
\vspace{1cm}
\section{Code Listings}
\begin{figure}[ph]
\begin{lstlisting}[language=grgen]
rule foldBinary(b:Binary<FirmNode>) : (Const)
{
	startBlock:StartBlock;
	b -df0:Dataflow-> c0:Const;
	b -df1:Dataflow-> c1:Const;
	hom(c0, c1);
	if{ df0.position==0 && df1.position==1; }

	alternative {
		foldAdd {
			if{ typeof(b)==Add; }

			modify {
				eval { c.value = c0.value + c1.value; }
			}
		}

		foldCmp {
			if{ typeof(b)==Cmp; }

			modify {
				eval {
					c.value = (
						cmp.relation == Relation::TRUE ||
						cmp.relation == Relation::GREATER
							&& c0.value >  c1.value ||
						cmp.relation == Relation::EQUAL
							&& c0.value == c1.value ||
						cmp.relation == Relation::GREATER_EQUAL
							&& c0.value >= c1.value ||
						cmp.relation == Relation::LESS
							&& c0.value <  c1.value ||
						cmp.relation == Relation::NOT_EQUAL
							&& c0.value != c1.value ||
						cmp.relation == Relation::LESS_EQUAL
							&& c0.value <= c1.value
					) ? 1 : 0;
				}
			}
		}

		...
	}

	modify {
		c:Const -startBlockEdge:Dataflow-> startBlock;
		eval { startBlockEdge.position = -1; }
		exec([relinkUser(b,c)] ;> deleteNode(b));
		return(c);
	}
}
\end{lstlisting}
	\caption{Rule for folding \texttt{Binary} nodes.}
	\label{fig:foldBinary.grg}
\end{figure}

\begin{figure}[ph]
\begin{lstlisting}[language=grgen]
rule foldPhi(phi:Phi<FirmNode>) : (Const)
{
	startBlock:StartBlock;
	phi -blockEdge:Dataflow-> block:Block;
	phi -e0:Dataflow-> c0:Const;

	iterated {
		phi -e:Dataflow-> c0;

		modify {
			delete(e);
		}
	}

	iterated {
		phi -e:Dataflow-> phi;

		modify {
			delete(e);
		}
	}

	negative {
		c0;
		block;
		phi -:Dataflow-> :FirmNode;
	}

	modify {
		delete(blockEdge, e0);

		c:Const<phi>;
		c -startBlockEdge:Dataflow-> startBlock;

		eval {
			c.value = c0.value;
			startBlockEdge.position = -1;
		}

		return(c);
	}
}
\end{lstlisting}
	\caption{Rule for folding \texttt{Phi} nodes.}
	\label{fig:foldPhi.grg}
\end{figure}

\begin{figure}[ph]
\begin{lstlisting}[language=grgen]
rule foldCond
{
	startBlock:StartBlock;
	cond:Cond -blockEdge:Dataflow->;
	cond -df0:Dataflow-> c0:Const;
	falseBlock:Block -falseEdge:False-> cond;
	trueBlock:Block -trueEdge:True-> cond;
	hom(falseBlock, trueBlock);

	alternative {
		TrueCond {
			if { c0.value == 1; }

			modify {
				delete(falseEdge);
				-jmpEdge:Controlflow<trueEdge>->;
			}
		}
		FalseCond {
			if { c0.value == 0; }

			modify {
				delete(trueEdge);
				-jmpEdge:Controlflow<falseEdge>->;
			}
		}
	}

	modify {
		delete(df0);
		jmp:Jmp<cond>;
	}
}
\end{lstlisting}
	\caption{Rule for folding \texttt{Cond} nodes.}
	\label{fig:foldCond.grg}
\end{figure}

\begin{figure}[ph]
\begin{lstlisting}[language=grgen]
rule removeUnreachableBlock
{
	block:Block\StartBlock;

	negative {
		block -cfgEdge:Controlflow->;
	}

	modify {
		delete(block);
	}
}
\end{lstlisting}
	\caption{Rule for removing unreachable \texttt{Block}s.}
	\label{fig:removeUnreachableBlock.grg}
\end{figure}

\begin{figure}[ph]
\begin{lstlisting}[language=grgen]
rule removeUnreachableNode
{
	n:FirmNode\Block;

	negative {
		n -blockEdge:Dataflow-> :Block;

		if { blockEdge.position == -1; }
	}

	modify {
		delete(n);
	}
}
\end{lstlisting}
	\caption{Rule for removing nodes without a \texttt{Block}.}
	\label{fig:removeUnreachableNode.grg}
\end{figure}

\begin{figure}[ph]
\begin{lstlisting}[language=grgen]
rule removeUnreachablePhiOperand
{
	phi:Phi -blockEdge:Dataflow-> block:Block;
	phi -operandEdge:FirmEdge-> operand:FirmNode;

	negative {
		block -cfgEdge:Controlflow->;

		if { cfgEdge.position == operandEdge.position; }
	}

	modify {
		delete(operandEdge);
	}
}
\end{lstlisting}
	\caption{Rule for removing nodes without a \texttt{Block}.}
	\label{fig:removeUnreachablePhiOperand.grg}
\end{figure}

\begin{figure}[ph]
\begin{lstlisting}[language=grgen]
rule foldAssociativeAndCommutative
{
	bin0:Binary;
	bin0 -:Dataflow-> block:Block;
	bin0 -binEdge:Dataflow-> bin1:typeof(bin0);
	bin0 -constEdge:Dataflow-> c0:Const;
	bin1 -:Dataflow-> operand:FirmNode\Block;
	bin1 -:Dataflow-> c1:Const;
	hom(c0, c1);

	if { bin0.associative && bin0.commutative; }

	modify {
		delete(binEdge, constEdge);

		bin:typeof(bin0);
		bin -blockEdge:Dataflow-> block;
		bin -left:Dataflow-> c0;
		bin -right:Dataflow-> c1;
		bin0 -binLeft:Dataflow-> operand;
		bin0 -binRight:Dataflow-> bin;

		eval {
			blockEdge.position = -1;
			left.position = 0;
			right.position = 1;
			binLeft.position = binEdge.position;
			binRight.position = constEdge.position;
		}
	}
}
\end{lstlisting}
	\caption{Rule for constant folding of associative and commutative operations.}
	\label{fig:foldAssociativeAndCommutative.grg}
\end{figure}

\begin{figure}[ph]
\begin{lstlisting}[language=grgen]
rule NormalizeConsts
{
	bin:Binary;
	bin -op0Edge:Dataflow-> op0:Const;
	bin -op1Edge:Dataflow-> op1:FirmNode\Const;

	if { bin.commutative && op0Edge.position == 0 && op1Edge.position == 1; }

	modify {
		eval {
			op0Edge.position = 1;
			op1Edge.position = 0;
		}
	}
}
\end{lstlisting}
	\caption{Rule for normalizing commutative operations with a \texttt{Const} operand.}
	\label{fig:normalizeConst.grg}
\end{figure}

\begin{figure}[ph]
\begin{lstlisting}[language=grgen]
rule ToTargetAddI
{
	a:Add -df:Dataflow-> c:Const;
	if { df.position==1; }

	modify {
		ta:TargetAddI<a>;
		delete(df);
		eval {
			ta.value = c.value;
		}
	}
}
\end{lstlisting}
	\caption{Rule for an \texttt{TargetAddI} operation.}
	\label{fig:iselAddI.grg}
\end{figure}

\begin{figure}[ph]
\begin{lstlisting}[language=grgen]
rule ToTargetLoad
{
	l:Load;
	modify {
		t:TargetLoad<l>;
		eval { t.volatile = l.volatile; }
	}
}
\end{lstlisting}
	\caption{Rule for an \texttt{TargetLoad} operation.}
	\label{fig:iselLoad.grg}
\end{figure}

\begin{figure}[ph]
\begin{lstlisting}[language=xgrs]
xgrs [NormalizeConst]
xgrs [ToTargetLoadI] | [ToTargetStoreI] | [ToTargetAddI] |\
   [ToTargetSubI]  | [ToTargetMulI]   | [ToTargetDivI] |\
   [ToTargetModI]  | [ToTargetAndI]   | [ToTargetOrI]  |\
   [ToTargetEorI]  | [ToTargetShlI]   | [ToTargetShrI] |\
   [ToTargetShrsI] | [ToTargetCmpI]

xgrs [ToTargetJmp]   | [ToTargetCond]  | [ToTargetLoad]   |\
   [ToTargetStore] | [ToTargetConst] | [ToTargetSymConst] |\
   [ToTargetNot]   | [ToTargetAdd]   | [ToTargetSub]      |\
   [ToTargetMul]   | [ToTargetDiv]   | [ToTargetMod]      |\
   [ToTargetAnd]   | [ToTargetOr]    | [ToTargetEor]      |\
   [ToTargetShl]   | [ToTargetShr]   | [ToTargetShrs]     |\
   [ToTargetCmp]
\end{lstlisting}
	\caption{Extended graph rewrite sequence for instruction selection.}
	\label{fig:isel.grs}
\end{figure}

\end{document}